# Protégé Effect for Behaviour Change: Does Teaching Digital Stress Solutions to Others Reduce One's Own?

Sameha Alshakhsi[1], Ala Yankouskaya[2], Dena Al-Thani[1], & Raian Ali[1]

[1]College of Science and Engineering, Hamad Bin Khalifa University, Doha, Qatar

[2]Department of Psychology, Bournemouth University, UK

**Abstract**. The protégé effect suggests that individuals learn a subject more effectively when they teach it to others. Can we therefore assume that when individuals with a problematic behaviour teach others about it and how to manage it, they become more likely to reduce and better manage that behaviour themselves? We address this question by evaluating a protégé based approach to reducing and managing digital stress (DS). Over three weeks, 137 participants with moderate-high DS were assigned to four groups. Two groups followed a protégé based approach: a passive group, given materials to teach, and an active group, given headlines and required to search for and prepare teaching content. Both groups completed three sessions, each focusing on one DS dimension: availability stress, approval anxiety, and fear of missing out. A digital literacy group received content and quizzes, alongside a control group. Outcomes included DS levels, problematic social media use, word of mouth, and issue involvement. Results showed no significant differences between groups. The findings highlight the difficulty of translating cognitive engagement into behavioural change. This study contributes empirical evidence to digital wellbeing literature by illustrating the challenges of enhancing digital wellbeing, potentially due to persistent digital habits and socially reinforced stressors.

## 1 Introduction

The widespread integration of digital technologies, particularly social media, has reshaped the way individuals communicate, engage, and maintain relationships. Today, over 70% of the global population owns a mobile phone, 83.7% of which are smartphones, making them a key gateway to digital life [1]. These devices real-time messaging, video calls, content sharing, and information access on a single platform. As a result, digital engagement has become pervasive, with more than 63% of the world's population using social media, averaging over two hours per day[2][3]. This engagement is more pronounced among adolescents and young adults, with nearly half of teenagers reporting being online almost constantly [4]. While this hyperconnected environment facilitates immediacy and continuous access, it also increases social expectations around availability, responsiveness, and social comparison, which form the emergence of digital stress[5].

Digital stress, a psychological phenomenon stemming from the integration of technology and social interactions, has emerged as a growing concern [6]. It is characterized by psychological strain resulting from the pressures of constant connectivity, approval-seeking, fear of missing out (FoMO), and connection overload. This stress is compounded by social pressure and expectations tied to social media use, such as the need to maintain a curated online presence, respond promptly to messages, and meet the demands of social media platforms[7,8]. Over time, these pressures can negatively impact one's wellbeing[9]. Prior research has linked digital stress to increased levels of stress, anxiety, burnout from social media use [8], mental fatigue [10,11], depressive symptoms [12,13], lower self-esteem[14]. These stressors can significantly affect an individual's overall wellbeing and mental health, underscoring the critical need for effective interventions to mitigate digital stress and promote better digital wellbeing.

One potential way to address such digital stressors is through interpersonal reflection and enhanced feeling of commitment to manage it. The protégé effect, also known as learning by teaching, suggests that individuals learn more effectively when they intend to teach others [15]. Teaching promotes deeper engagement with the material and encourages reflection. In the context of digital stress, we posit that when individuals are asked to teach others about the concepts of digital stress and techniques for managing it, they will engage in reflection on these techniques and consequently relate them to their own experiences, which help internalizing the strategies they are promoting. We hypothesize that when individuals explain to others the different causes of digital stress and coping techniques, they are simultaneously reinforcing their own understanding and developing an internal commitment to apply these strategies themselves. This aligns with Cialdini's principle of commitment and consistency, which suggests that once individuals commit to a behavior, they are more likely to follow through due to a desire to appear consistent with their commitment[16]. Consequently, this intervention may support reflection on the causes of digital stress and enhance motivation to apply coping strategies, particularly those related to managing social expectations, more effectively.

Given the above literature, this study aims to design and evaluate a socio-technical intervention based on the protégé effect to help manage digital stress. In our study, we followed a mixed-factorial design where participants were divided into four groups: (1) a passive protégé group, where participants receive reading materials (covering types of digital stress and coping strategies) and are tasked to use it for preparing materials to teach others about digital stress and its coping strategies, (2) an active protégé group where participants receive reading materials (covering types of digital stress) and are tasked to search for strategies to manage digital stress and prepare teaching material accordingly, (3) a digital literacy group, where participants only read the provided materials about digital stress (covering types of digital stress and coping strategies) and answer quizzes, and (4) a control group with no intervention.

Digital stress and problematic social media use (PSMU) are related constructs. Digital stressors including constant pressure to maintain online availability, seeking social approval, and FoMO, associated with PSMU patterns, including excessive checking, and social media use [17] [18][19]. Furthermore, research revealed that digital stressors can play a mediator role between PSMU and adverse mental health outcomes [20] [21]. In light of this, addressing digital stress is assumed to indirectly reduce the level of PSMU. This notion aligns with principles from inoculation theory, which suggest that building psychological resistance in one domain may confer protective effects across a related domain [22]. The protégé effect, similar to inoculation theory, involves cognitive engagement with content, which may strengthen individuals' attitudes and prompt reflective thinking. Therefore, in our study, while our intervention aims to reduce level of digital stress, it is hypothesized to yield secondary benefits by reducing the level of PSMU.

To further evaluate the effectiveness of the intervention and its influence on participants' attitudes toward managing digital stress, two additional constructs are considered: issue involvement and word-of-mouth (WOM) intentions. Issue involvement refers to the level an individual perceives a topic as personally important or relevant [23]. When participants perceive the intervention content as relevant to their own experiences, they are more likely to engage in deeper cognitive processing, which can enhance comprehension and reflection [24]. In the context of and intervention based on the protégé effect to manage digital stress, the act of preparing to teach others and engaging with the topic is expected to increase individuals' sense of relevance and motivation to share with others. This aligns with the WOM spread, which posits that individuals are likely to talk and spread information that positively perceived [25]. Furthermore, drawing on inoculation theory, engaging participants in an issue that is relevant to participants encourages reflection [26]. Therefore, while our intervention which involves teaching and potentially reflection may or may not reduce digital stress, it would foster higher levels of issue involvement and increase the likelihood of positive WOM about digital stress management.

Building on the above literature, this research addresses the following questions:

**RQ1**: Can protégé-based intervention (passive and active types) reduce overall digital stress and its components (availability demand stress, approval anxiety, and FoMO)?

**RQ2**: Can protégé-based intervention (passive and active types) reduce problematic social media use?

**RQ3**: Do the interventions (passive protégé-based, active protégé-based, digital literacy training) increase levels of issue involvement, i.e. involvement in digital stress management?

**RQ4**: Do the interventions (passive protégé-based, active protégé-based, digital literacy training) affect participants' intentions to talk with others about managing digital stress (word-of-mouth (WOM)), by increasing the positive WOM and decreasing the negative WOM?

## 2 Methods

### 2.1 Theoretical underpinnings

***Digital stress*** the multifaceted nature of digital stress encompasses various stressors that contribute to overall distress. These include availability demands, approval anxiety, fear of missing out (FoMO), connection load, and online vigilance. Such stressors are often exacerbated by social pressures and expectations that reinforce the need for constant availability, social validation, and continuous engagement. To better understand these dynamics, we examine these dimensions of digital stress.

***Availability demands stress*** refers to the pressure individuals feel to remain constantly accessible for online interactions. Social media, while providing contact opportunities that can improve relationships and social bonding[27], also introduces distress due to the perceived expectations that individuals face to stay constantly connected and respond promptly to messages, comments, or notifications. These expectations can be driven by social norms, social cues, such as "seen" or "read" notifications, and implicit or explicit expectations from others, creating a pervasive sense of obligation to respond immediately to maintain relationships [28] [29]. This type of stress can be understood through several theoretical lenses. Social Identity Theory (SIT) [30], which suggests that individuals derive a significant part of their self-concept from their social groups. As such, they may internalize the expectation of constant availability to reinforce their group membership and avoid exclusion. Failure to meet these expectations can threaten their social identity. Similarly, Cialdini's principles on social influence and conformity[31] explains how external pressures push individuals to conform to social norms, which can include being constant connectivity, to avoid disapproval or rejection. Factors such as the need to belong and the fear of exclusion drive individuals to conform and compel to social norms such as digital availability expectations, making this demand a source of stress [31].

***Approval Anxiety*** is stress related to concerns about how others, particularly peers, will react to one's social media posts and profiles, driven by the desire to be seen positively and validated online. This anxiety arises from the social pressure to craft an appealing profile and maintain social standing which can be facilitated through social media feedback features such as likes, comments, and shares. Drawing on Impression Management Theory [32][33] [34], individuals strategically curate their online persona to control how they are perceived, motivated by a need for social approval and validation. Social pressure amplifies this stress, as individuals feel compelled to conform to the perceived standards set by others. The public nature of social media interactions intensifies this anxiety, as individuals worry about negative feedback or lower engagement, a phenomenon known as Fear of Negative Evaluation (FNE)[35][36][37]. Additionally, this phenomenon can be fuelled by Social Comparison Theory[38], which explains how individuals tend to compare themselves to others. Engaging in upward social comparisons, that occur when comparing oneself to those perceived as more successful or popular, can lead to feelings of inadequacy and lower self-worth[39].

***Fear of Missing Out (FoMO)*** is the anxiety and fear that others are engaging in rewarding experiences without one's participation [40]. This is often exacerbated by the constant updates and high visibility into others' lives provided by social media [41]. This feeling is further heightened by social pressure, as people use digital media not only to stay connected but also to keep track of others' activities. Recent work by Alutaybi et al. [41] expands on this by identifying several distinct experiences of FoMO, such as anxiety when users do not receive expected feedback (likes, comments), feeling obliged to be connected due to social pressures including belonging and impression management, or when they are unable to participate in online interactions due to technical or time constraints. These findings highlight that FoMO is not just about missing out, but also about navigating social expectations around being engaged and responding to social pressure. FoMO leads to compulsive social media use, where users constantly check their social media, refresh their feeds or check stories to avoid feeling left out [41][42]. This is driven by psychological needs such as belonging, popularity, and fear of exclusion, motivating individuals to remain continually engaged and updated on others' experiences [41]. Studies have shown that FoMO is associated with lower mood, reduced life satisfaction, and problematic social media use, further contributing to digital stress [40] [43].

***Connection Overload***. Stress from the overwhelming number of digital inputs, such as notifications, messages, and updates that individuals receive from various platforms[6]. This constant influx of information can lead to feelings of being overwhelmed and anxious, as individuals struggle to keep up with the demands of staying connected and engaged across multiple channels, leading to a sense of information overload and loss of control[44][45]. The pressure to respond quickly, coupled with the fear of missing important communications, amplifies this stress [8]. Furthermore, research has shown that connection overload mediates the relationship between digital multitasking, such as checking one's phone during face-to-face interactions, and increased depressive symptoms [46].

***Online vigilance.*** Online vigilance is cognitive preoccupation with online connections, and mental readiness to check updates and react to online connections, even if it means periodizing online over offline activities [47]. This hyperalert can cause mental fatigue, which can reduce emotion regulations [10]. In contrast to the socially driven dimensions of digital stress, such as availability stress, approval anxiety, and FOMO, connection overload and online vigilance are linked to the design of social media and mobile platforms[48]. These stressors are shaped by digital features that encourage constant connectivity and user responsiveness, such as notifications that can heighten perceived connection overload and online vigilance [7][47].

***Protégé Effect*** The act of teaching necessitates a deeper engagement with the content: selecting key information, organizing it meaningfully, and reflecting on how it integrates with existing knowledge. Research have also shown that protégé effect using teachable agents (e.g. computer based or game like character) may involve a perceived sense of responsibility toward the learner, which enhances efficiency and experience of learning outcome[49][50]. Drawing on the teaching expectancy effect [51] [52], even preparing to teach (without actual teaching) can also be effective. It enhances cognitive processing by encouraging the identification of key information and organizing them in structure way to present them. Reflective practice further supports the importance of commitment in learning, as individuals who learn and reflect on are more likely to internalize and apply what they learn, reinforcing their understanding and improving their action [53]. While the protégé effect has been mostly studied in academic domains (e.g., mathematics, programming), evidence for its role in behavior change is limited. Related research shows that taking on teaching roles can shape one's own actions. For instance, in a workplace context, a study found that mentoring others led to improved performance and prosocial behavior among mentors, driven by increased psychological meaning and a stronger sense of responsibility [54]. This raises the question of whether behaviors, like managing digital stressors, can also be shaped through teaching. Though more empirical research is needed, these mechanisms suggest that the protégé effect provides a promising pathway not only for enhancing knowledge but also for supporting behavioral change through reflection, responsibility, and internalized commitment. Together, these mechanisms suggest that the protégé effect provides a promising approach for interventions to helping individuals reflect on and internalize strategies for managing social pressures and expectations in digital environments.

***Passive vs. Active Protégé-based Interventions*** According to a recent review by Rieh et al. [55], engaging in search-based learning activates cognitive monitoring and allows for personalized learning, making it a meaningful form of active learning. Additionally, parallels can be drawn from inoculation theory approaches, where interventions often compare passive exposure to refutational messages with active generation of counterarguments in response to persuasive content[56][57]. Although studies show mixed results between these approaches, the differences have been attributed to the cognitive engagement and effort required during active inoculation[58]. Following this logic, the current study explores whether passive and active protégé-based interventions will yield different outcomes in managing digital stress.

### 2.2 Participants

A total of 137 participants were recruited online from Middle East background. The sample comprised 59.12% females. Participants were recruited through the online platform Prolific (http://www.prolific.com) and were required to meet eligibility criteria including being between 18 and 35 years old, identifying with Middle Eastern cultural norms, being active users of social media, exhibiting moderate to high levels of digital stress, and having access to a laptop with PowerPoint or Google Slides capabilities so they can prepare teaching material if requested. Eligibility was assessed via a pre-screening survey, and only individuals meeting all criteria were invited to the main study. To ensure comparability across conditions, efforts were made to balance gender and baseline digital stress levels across groups.

Data collection occurred between January 2025 and mid-April 2025. The study was approved by the Institutional Review Board (IRB) of the first author's institution, ensuring participants provided informed consent and were informed that they could withdraw from the study at any time. Informed consent was obtained electronically at the start of the study. To ensure data quality, participants who failed attention checks or completed the survey in a speedy manner were removed. Participants received monetary compensation for each completed phase, including the pre- and post-surveys and the intervention sessions.

### 2.3 Study Design

The present study employs a 4 (group) × 2 (time: pre, post) mixed design to evaluate the impact of an intervention based on the protégé effect or 'learn by teaching' on manage digital stress. Participants were invited to the study through Prolific and were contacted during the experiment via Prolific's messaging system. They engaged with the intervention materials and related questions, which were delivered through SurveyMonkey and Google Docs. Participants in the intervention conditions engaged with a series of materials specifically developed to address three core components of digital stress (availability demand stress, approval anxiety, and FoMO), particularly those stemming from social expectations in online environments. The intervention was structured into three distinct sessions, with each session dedicated to one of these digital stress components. The materials were developed based on relevant literature on digital stress. Study materials, including surveys, intervention content, and templates, are provided in the Supplementary Materials section.

To guide task completion, participants were provided with a PowerPoint template (see Figure 1), which included space for two slides per session. Each slide prompted participants to develop one digital stress scenario related to the session's topic and to propose coping strategies in response. Thus, participants were expected to create two unique scenario-strategy pairs per session. This template served as a structured framework outlining required elements, standardizing task expectations, simplify the process and reducing ambiguity across participants. To further enhance understanding, a worked example was provided illustrating how to complete the slides appropriately, as shown in Figure 2. Participants were informed that feedback and clarifying questions would be provided by the learner to simulate a teaching context.

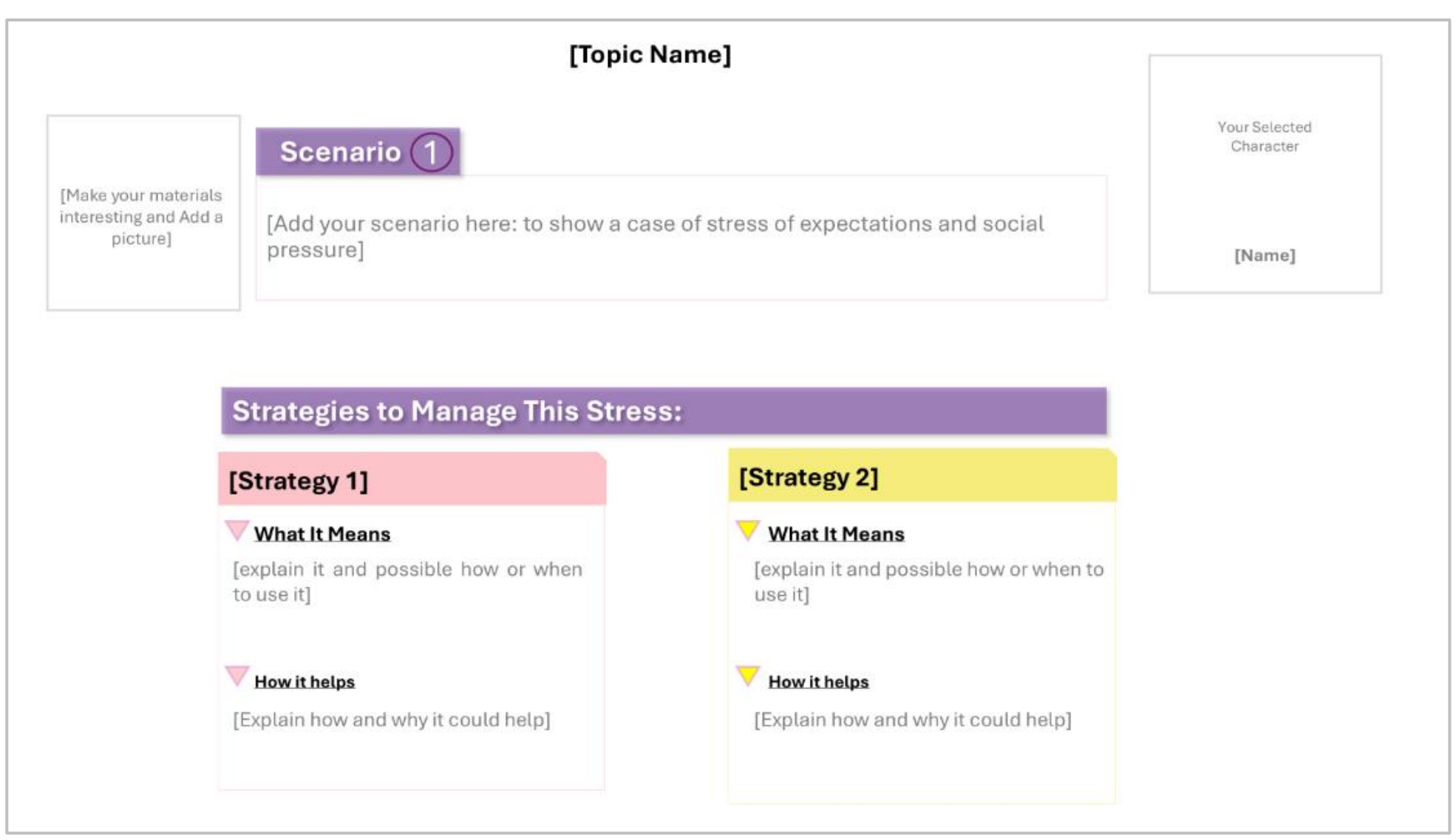

*Figure 1 Template provided to participants for preparing teaching materials*

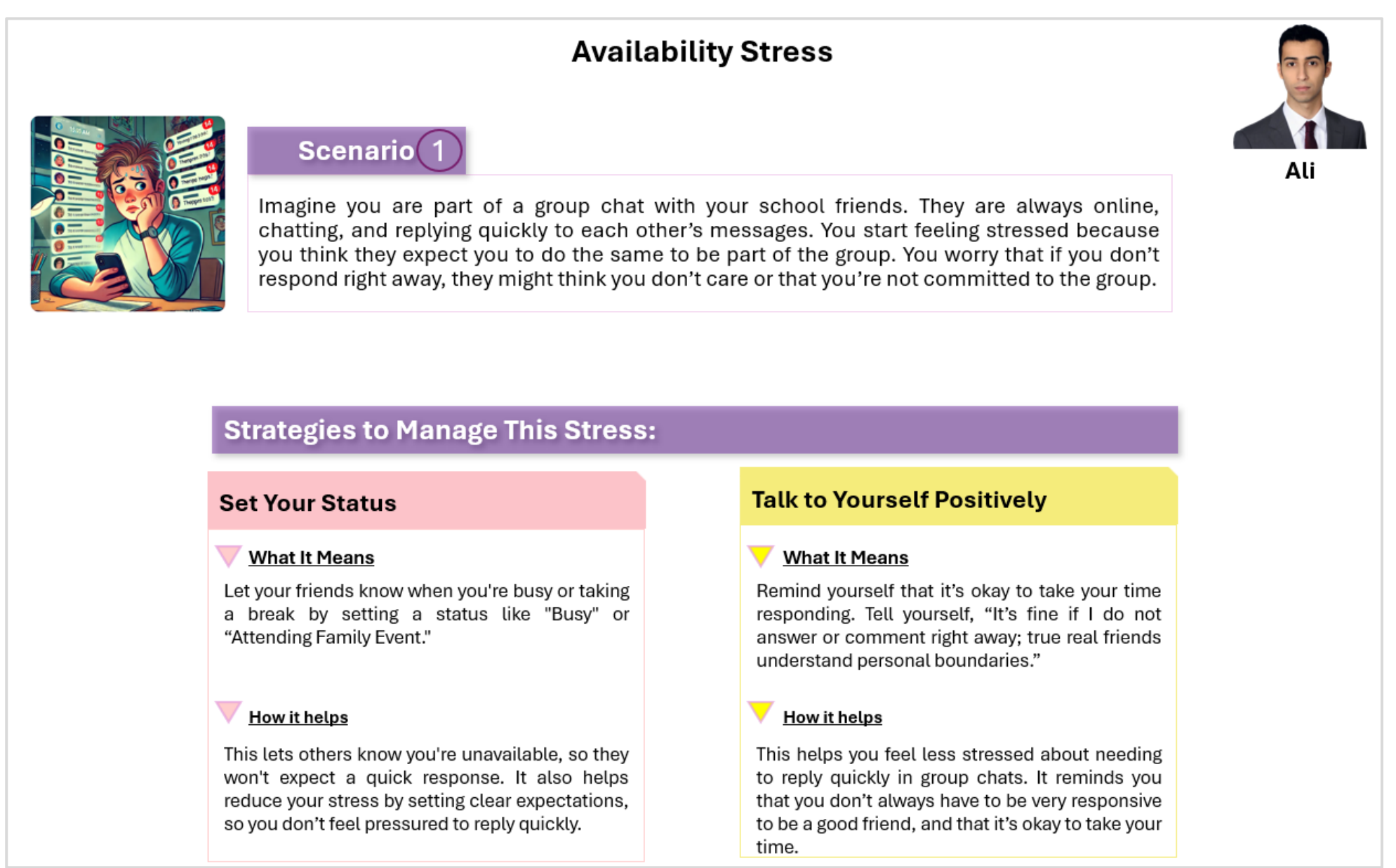

*Figure 2 Example of completed teaching materials shown to participants (photos are AI generated)*

The minimum sample size was calculated using Statistical Power Analysis (G*Power) software, calculated the effect size for a repeated measures ANOVA within-between interaction design. Considering statistical power ($1-\beta$) of 80%, a significance level ($\alpha$) of .05, and a small effect size (f) of 0.15, the required sample size was estimated to be 128 participants. To account for potential dropout or incomplete data, the target sample was increased by 20% as recommended in literature [59], yielding a recruitment goal of 160 participants. The adjusted sample size was calculated using formula N1 = n / (1 - d), where N1 is the adjusted sample size, n is the required sample size, and d represents the estimated dropout rate [60]. The final sample of 137 participants met the criteria for adequate statistical power.

**Experimental Procedure:**

The procedural flow of the study was divided into three main phases: baseline assessment, the intervention period, and the post-intervention assessment. This structure was intended to capture changes in attitudes and perceived digital stress across conditions.

**Pre-screening- Eligibility assessment.** At this stage, participants were introduced to the study's purpose, the multi-phase design, and what would be expected of them. Eligibility criteria were then assessed to determine which individuals would be invited to the next stage of the study. A key criterion was the self-reported frequency of experiencing digital stress components, only those who reported experiencing digital stress at a moderate to high level were invited to continue. To ensure that participants held a consistent understanding of the constructs being measured, brief definitions of digital stress and its core components were provided before the screening items. These definitions are presented in Table 1.

Table 1. Definitions of digital stress components utilized in the study as provided to the Participants

| |
|---|
| **Digital stress** refers to the various kinds of pressure and stress people experience due to their interactions with others online such as on social media, messaging apps, and emails. Social pressure, e.g. to respond and be online, may contribute to digital stress. Digital stress has several factors including: |
| **Availability Stress**: the pressure to always be online and respond immediately to messages, notifications, or posts, often driven by the expectation that others will notice and judge your relationship with them by how quickly you reply. |
| **Approval Anxiety**: the anxiety or worry about how people will react to your posts, photos, or messages on social media, e.g. through likes, comments, or positive feedback. It is a human need to be approved and appreciated by their peers. |
| **FoMO** (the fear of missing out) refers to the fear of not being able to know what is happening on social media and participate in it and take opportunities. |

To improve comprehension and draw participants' attention to key constructs, the digital stress and its components (Availability Stress, Approval Anxiety, and FoMO) were visually emphasized throughout the study materials. This included consistent use of bold text and color formatting across definitions, survey questions, and intervention content.

***Phase 1 - Pre-Intervention Survey***. Prior intervention, participants completed a baseline survey assessing digital stress, PSMU, issue involvement, and word-of-mouth intention. Attention checks were embedded within the survey, and participants who failed these checks were excluded from further participation.

***Phase 2 - Intervention Phase***. Participants were assigned to one of four groups: passive protégé-based group, active protégé-based group, digital literacy, and no-intervention control group.

To provide a basis for this group assignment, the intervention was designed to examine whether different types of learning engagement, particularly through the protégé effect, might influence reflection and internalization. Accordingly, the study included two protégé conditions: passive and active. In the passive group, participants were provided with complete content and instructed to prepare teaching materials based on this content, relying on the teaching expectancy effect, which has been shown to promote learning and motivation through the act of explaining and organizing material for others. In contrast, the active group was required to search for coping strategies independently, drawing on the premise that searching is a learning process by itself that could enhance reflective thinking and critical evaluation.

To simulate a teaching experience, participants in both protégé groups were asked to select a learner character from a predefined list and prepare instructional content aimed at teaching this character how to cope with digital stress. This setup was intended to simulate a real teaching scenario and foster a sense of responsibility toward the learner, consistent with prior research on teachable agents.

The study groups received the following intervention procedures:

- **Group 1: Passive Protégé-based Group**. Participants received study materials covering three types of digital stress (availability stress, approval anxiety, and FoMO) and corresponding coping strategies. They were instructed to use this content to prepare two slides per session for a hypothetical learner. Each slide included a scenario and suggested strategies to manage the specific form of stress. Participants were provided with feedback and clarifying questions as required. A structured slide template and example were provided.
- **Group 2: Active Protégé-based Group**. Participants received partial materials (descriptions of stress types only), with instructions to independently search for appropriate coping strategies through online search. Recommended resources were provided to support this process. As with Group 1, participants developed two slides per session aimed at teaching a learner.
- **Group 3: Literacy Group**. Participants received the full materials (similar to Group 1) and were asked to read the content carefully. No slide creation was required. Instead, comprehension questions followed each reading task to ensure engagement.
- **Group 4: Control Group**. Participants did not receive any intervention materials or tasks between the pre- and post-survey.

Participants in Groups 1 and 2 completed three sessions over a 17-day period as shown in Figure 3.

Each session required participants to submit their slides within 24 hours. The research team responded to each submission with feedback and follow-up questions, framed as inquiries from the learner character. This dialogic exchange was designed to stimulate teaching context and potentially deeper reflection.

To ensure data quality and genuine engagement, quality assurance procedures were implemented throughout the intervention. Submissions that showed signs of duplication, minimal effort, or limited understanding (for example, generic responses or examples unrelated to digital contexts such as stress in offline settings) were returned to participants for revision. These checks were put in place to ensure participants engaged thoughtfully with the material and completed the learning tasks as intended.

***Phase 3 - Post-Intervention Survey.*** After the final session, all participants completed a post-survey with the same set of measures as the pre-survey. The aim was to evaluate changes in digital stress, PSMU, issue involvement, and word-of-mouth intentions.

**Study Timeframe.** The intervention phase spanned a 17-day period, during which participants completed three structured sessions on Days 4, 7, and 10. Time was intentionally spaced between sessions to allow participants to reflect on the material, receive and engage with feedback. The post-intervention survey was administered on Day 17, seven days after the final session, to provide sufficient time that would allow the cognitive and attitudinal effects of the intervention to consolidate. This timing is informed by inoculation theory research, which suggests that a delay between intervention exposure and outcome measurement is needed is needed for the intervention to take effect [61].

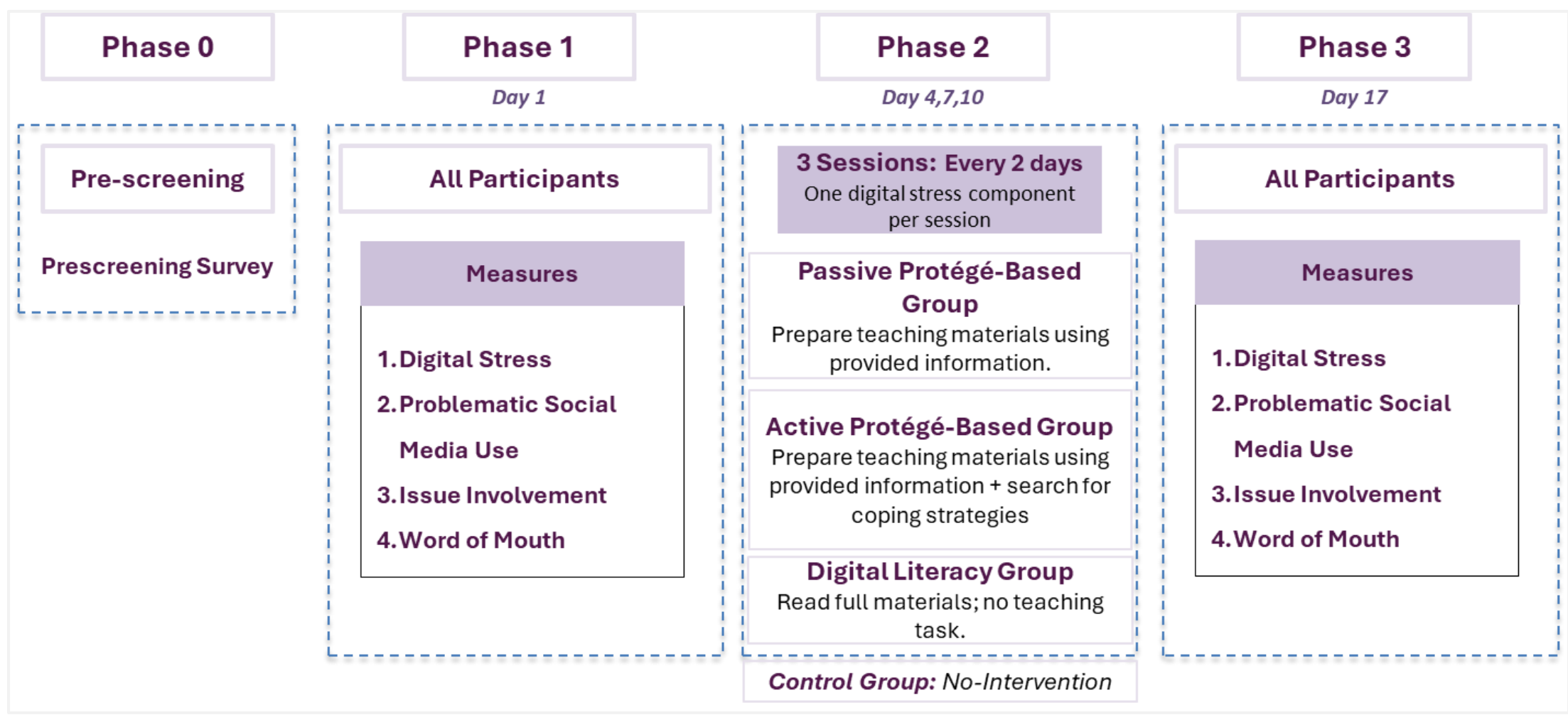

Figure 3 *Study Flow of the Protégé-based Intervention*

**Measures**:

**Digital Stress**

Digital stress was assessed using the Multidimensional Digital Stress Scale (DSS) developed by Hall et al. [48]. The DSS consists of 24 items capturing stress experiences related to digital technology use, particularly social media. Each item is rated on a 5-point Likert scale ranging from 1 ("Never") to 5 ("Always"), higher scores reflect greater perceived digital stress. The scale is composed of five dimensions: availability stress (5 items; e.g., "My friends expect me to be constantly available online"), approval anxiety (6 items; e.g., “I worry about how people react to my posts”), fear of missing out or FoMO (4 items; e.g., “I fear my friends are having more rewarding experiences than I am”), connection overload (4 items; e.g., “I have to check too many notifications”), and online vigilance (5 items; e.g., “I must have my phone with me to know what is going on”).

In the current study, digital stress scores were calculated as the sum of all items, while each subscale (component) score was computed by aggregating its relevant items. The scale demonstrated good reliability, with Cronbach’s alpha values of $\alpha = .86$ at pre-intervention and $\alpha = .90$ at post-intervention.

**Problematic social media use (PSMU)**

To assess PSMU, the study utilized the Social Media Disorder (SMD) Scale [62]. This instrument includes nine items, each corresponding to a distinct behavioral symptom of social media addiction: preoccupation, persistence, tolerance, withdrawal, displacement, escape, interpersonal problems, deception, and conflict. An example item from the scale is: “… regularly found that you can't think of anything else but the moment that you will be able to use social media again?”. Participants rated each item using a 5-point Likert scale ranging from 1 ("Never") to 5 ("Always"). Scores were summed across all items to yield a total PSMU score, with higher scores reflecting greater levels of problematic use.

The scale showed good internal consistency in this study, as evidenced by Cronbach’s alpha coefficients of $\boldsymbol{\alpha} = .83$ at pre-intervention and $\boldsymbol{\alpha} = .85$ at post-intervention, indicating a good reliability in this sample.

**Issue involvement**

Participants’ level of issue involvement was measured using an abbreviated version of the Personal Involvement Inventory (PII), originally developed by Zaichkowsky [23]. This scale is frequently utilized in inoculation and persuasion research to gauge individuals’ personal relevance to a topic.

In this study, the scale was adapted to assess participants’ issue involvement in managing digital stress, particularly in relation to their social media use. Participants responded to six bipolar adjective pairs: unimportant–important, irrelevant–relevant, of no concern–of concern to me, meaningless–meaningful, does not matter–matters to me, and insignificant–significant. Participants rated each pair in relation to the topic of managing digital stress, using a 7-point Likert scale. The question began with the prompt “The next set of items assesses the importance you place on managing digital stress. Thinking about my social media use,” followed by items for each bipolar adjective pair. For example: “Managing digital stress is…” rated on a scale from 1 (unimportant) to 7 (important).

The scale demonstrated good internal consistency in this study. Cronbach’s alpha coefficients were $\alpha = .93$ before the intervention and $\alpha = .93$ after the intervention, indicating good reliability in both assessments.

**Word-of-mouth (WOM):**

WOM intentions were assessed using a unidimensional scale that captured participants' likelihood of engaging in interpersonal discussions about the topic [63]. Participants rated three statements on a 7-point Likert scale ranging from 1 ("Strongly disagree") to 7 ("Strongly agree"), indicating their willingness to talk about, recommend, or encourage digital stress management strategies.

Example items included: "I will say positive things about managing digital stress to others," and "I will try to recommend strategies to manage digital stress to someone who seeks my advice." These items were used to assess participants' willingness to talk about the topic after engaging with the intervention.

## 2.4 Data analysis

A 2 × 4 mixed-design repeated measures ANOVA was used as the main analysis to test the effect of Protégé-based intervention on digital stress (total score of digital stress), as well as for each component of digital stress (availability demand stress, approval anxiety, FoMO, connection Overload, and Online Vigilance). A separate analysis will be conducted to assess potential changes in problematic social media use (PSMU). The analysis includes two time points (within-subjects: pre- and post-intervention) and four groups (between-subjects). Data normality was checked using skewness and kurtosis values [64]. The scales assessing digital stress, and each component, and PSMU exhibited skewness and kurtosis between ±2, indicating an approximately normal distribution. Data were analysed using JASP version 0.19.3 [65].

# 3 Results

## 3.1 Demographics

Table 2 presents a summary of the demographic characteristics of participants. Descriptive statistics for the main study variables are provided in Table 3, including digital stress, its three components (availability demand stress, approval anxiety, and FoMO), as well as problematic social media use, issue involvement, and word-of-mouth (WOM).

Table 2. Demographic characteristics of participants

| **Variables** | **N=137** | | | | |
|---|---|---|---|---|---|
| **Gender (%)** | Total | Passive PE Group | Active PE Group | Digital Literacy Group | Control Group |
| Male | 56 (40.88%) | 14 (42.42%) | 15 (42.86%) | 15 (41.67%) | 12 (36.36%) |
| Female | 81 (59.12%) | 19 (57.58%) | 20 (57.14%) | 21 (58.33%) | 21 (63.64%) |
| **Age** | | | | | |
| M (SD) | 26.82 (4.28) | 27.42 (4.37) | 26.97 (3.94) | 25.86 (4.46) | 27.09 (4.34) |
| Range | 18 - 35 | 21 - 35 | 20 - 34 | 18 - 35 | 19 - 35 |

Table 3. Descriptive statistics for study variables

| Measure | Passive PE Group | | Active PE Group | | Digital Literacy Group | | Control Group | |
|---|---|---|---|---|---|---|---|---|
| | Pre | Post | Pre | Post | Pre | Post | Pre | Post |
| Availability demand stress | 25.03 (6.45) | 25.06 (6.10) | 24.54 (8.17) | 23.97 (6.35) | 25.00 (7.73) | 22.06 (7.91) | 22.55 (9.00) | 19.70 (9.23) |
| Approval Anxiety | 43.94 (9.61) | 38.49 (12.36) | 44.40 (8.84) | 40.23 (11.22) | 44.67 (10.22) | 40.17 (9.53) | 46.15 (10.01) | 39.56 (11.77) |
| FoMO | 24.97 (7.80) | 23.61 (8.35) | 24.29 (8.01) | 22.80 (8.38) | 24.28 (7.28) | 23.92 (6.83) | 26.18 (6.73) | 23.85 (7.60) |
| Connection Overload | 40.18 (10.29) | 33.39 (11.64) | 42.80 (9.69) | 39.40 (10.38) | 42.42 (7.11) | 36.08 (10.79) | 40.36 (11.72) | 36.64 (12.17) |
| Online Vigilance | 31.12 (6.94) | 27.15 (7.01) | 31.08 (6.55) | 27.29 (6.66) | 31.33 (6.15) | 27.61 (7.58) | 30.06 (6.09) | 26.79 (7.75) |
| Digital stress | 165.24 (26.88) | 147.70 (30.47) | 167.11 (25.31) | 153.69 (30.59) | 167.69 (24.61) | 149.83 (26.26) | 165.30 (24.15) | 146.55 (32.17) |
| Problematic social media use | 46.82 (15.05) | 37.76 (13.94) | 48.26 (13.50) | 39.00 (13.40) | 47.36 (13.07) | 41.17 (12.04) | 46.52 (14.98) | 41.94 (17.52) |
| Issue involvement | 36.79 (4.94) | 37.06 (4.66) | 36.80 (4.19) | 36.83 (5.01) | 37.28 (5.40) | 37.89 (3.96) | 35.61 (5.78) | 35.94 (5.12) |
| Word-of-mouth | 17.27 (3.38) | 17.42 (2.44) | 17.80 (2.83) | 18.43 (2.19) | 17.61 (2.69) | 18.56 (1.90) | 17.52 (3.00) | 16.64 (2.51) |

### 3.2 Protégé-based intervention for managing digital stress

A series of repeated-measures ANOVAs were conducted to examine to examine changes in digital stress and its components, including availability demand stress, approval anxiety, fear of missing out (FoMO), connection overload, online vigilance, and overall digital stress, across time (pre- and post-intervention).

The Protégé-based intervention had a significant within-subject main effect on all six components of digital stress and overall digital stress (Table 4) and confirmed by post hoc comparisons (Table 5). This indicates that all participants experienced reductions in digital stress following the intervention. However, no significant interaction or main effects of group were observed across any of the digital stress measures, suggesting that the change over time was consistent regardless of group and that the changes reflect general effects of participation over time rather than differential responses to the specific group conditions.

Table 4. Effect on Availability demand stress, Approval Anxiety, FoMO, and overall digital stress

| Measure | **Within-subject main effect** | **Interaction effect** | **Main Effect of Group** |
|---|---|---|---|
| Availability demand stress | F(1,133)= 6.898, p = **.010**, $\eta p^2$ = 0.049 | F(3,133)= 1.617, p = .188, $\eta p^2$ = 0.035 | F(3,133)= 2.034, p = .112, $\eta p^2$ = 0.044 |
| Approval Anxiety | F(1,133)= 30.444, p < **.001**, $\eta p^2$ = 0.186 | F(3,133)= 0.329, p = .804, $\eta p^2$ = 0.007 | F(3,133)= 0.204, p = .894, $\eta p^2$ = 0.005 |
| FoMO | F(1,133)= 5.954, p = **.016**, $\eta p^2$ = 0.043 | F(3,133)= 0.512, p = .675, $\eta p^2$ = 0.011 | F(3,133)= 0.265, p = .851, $\eta p^2$ = 0.006 |
| Connection Overload | F(1,133)= 30.134, p **< .001**, $\eta p^2$ = 0.185 | F(3,133)= 0.894, p = .446, $\eta p^2$ = 0.020 | F(3,133)= 1.325, p = .269, $\eta p^2$ = 0.029 |
| Online Vigilance | F(1,133)= 46.771, p **< .001**, $\eta p^2$ = 0.260 | F(3,133)= 0.073, p =.974, $\eta p^2$ = 0.002 | F(3,133)= 0.181, p = .909, $\eta p^2$ = 0.004 |
| Digital stress | F(1,133)= 44.051, p **< .001**, $\eta p^2$ = 0.249 | F(3,133)= 0.219, p = .883, $\eta p^2$ = 0.005 | F(3,133)= 0.271, p = .846, $\eta p^2$ =0.006 |

Table 5. Post Hoc Comparison - Pre-intervention vs Post-intervention

| **Measure** | **Mean Difference** | **t(df)** | **Effect size (Cohen's d)** | **$p_{bonf}$** |
|---|---|---|---|---|
| Availability demand stress | 1.58 | 2.63 (133) | 0.21 | .010 |
| Approval Anxiety | 5.18 | 5.52 (133) | 0.49 | <.001 |
| FoMO | 1.39 | 2.44 (133) | 0.18 | .016 |
| Connection Overload | 5.06 | 5.49 (133) | 0.48 | <.001 |
| Online Vigilance | 3.69 | 6.84 (133) | 0.54 | <.001 |
| Digital stress | 16.90 | 6.64 (133) | 0.61 | <.001 |

Taken together, post hoc comparisons showed that the intervention produced small effects on availability demand stress (d = 0.21, 95% CI [0.05, 0.36]) and FoMO (d = 0.18, 95% CI [0.03, 0.33]). Moderate effects were observed for approval anxiety (d = 0.49, 95% CI [0.31, 0.68]) and connection overload (d = 0.48, 95% CI [0.30, 0.66]). The strongest effects emerged for online vigilance (d = 0.54, 95% CI [0.37, 0.71]) and overall digital stress (d = 0.61, 95% CI [0.41, 0.81]), indicating that the intervention had its greatest impact on reducing vigilance and general digital stress, while effects on availability demands and FoMO were comparatively small.

### 3.3 Does Protégé-based intervention for managing digital stress affect problematic social media use?

As shown in Table 6, there was a significant within-subject main effect on problematic social media use, indicating a reduction after the intervention. This reduction was further confirmed by post hoc analysis (MD = 7.27, t(133)=7.19, p<.001, Cohen's d = 0.51, 95%CI [0.36, 0.67]. Consistent with the findings for digital stress, no significant interaction or group effects were found. This indicate that the type of intervention (passive protégé-based, active protégé-based, digital literacy, or control) did not differentially impact the reduction in problematic social media use.

Table 6. Effect on problematic social media use

| Measure | **Within-subject main effect** | **Interaction effect** | **Main Effect of Group** |
|---|---|---|---|
| Problematic social media use | F(1,133)= 51.730, p **< .001**, $\eta p^2$ = 0.280 | F(3,133)= 1.252, p = 0.294, $\eta p^2$ = 0.027 | F(3,133)= 0.271, p = .846, $\eta p^2$ = 0.006 |

### 3.4 Does Protégé-based intervention for managing digital stress affect Issue involvement and WOM?

As presented in Table 7, issue involvement demonstrated significant within-subject main effects, indicating increased engagement after the intervention. However, there were no significant interaction effects or group differences again suggesting uniform effects across groups.

For word-of-mouth (WOM), the analysis showed that the main effect of time on WOM intentions was not significant, suggesting no overall change in WOM across time points. However, the interaction between time and group was significant, indicating that the pattern of change in WOM intentions over time differed by group. However, post hoc comparisons revealed no significant within-group changes in WOM between pre and post measurements across all groups (MD = -0.21, t(133)=-0.88, p=.380).

Table 7. Effect on Issue involvement and Word-of-mouth (WOM)

| Measure | **Within-subject main effect** | **Interaction effect** | **Main Effect of Group** |
|---|---|---|---|
| Issue involvement | F(1,133)= 0.621, p =.432, $\eta p^2$ = 0.005 | F(3,133)= 0.095, p =.963, $\eta p^2$ = 0.002 | F(3,133)= 1.023, p =.385, $\eta p^2$ =0.023 |
| Word-of-mouth (WOM) | F(1,133)= 0.776, p =.380, $\eta p^2$ =0.006 | F(3,133)= 2.738, p **=.046**, $\eta p^2$ = 0.058 | F(3,133)=1.869, p =.138, $\eta p^2$ = 0.040 |

## 4 Discussion

The present study proposed and evaluated a socio-technical intervention based on the protégé effect to reduce digital stress and its key components, as well as its influence on problematic social media use (PSMU), issue involvement, and word-of-mouth (WOM) intentions. Our participants reported significant reductions in digital stress and its components from pre- to post-assessment. In the intervention conditions, these improvements are consistent with the way the programme required participants to engage actively with learning strategies. Preparing materials for a learner, responding to scenarios, and revising work after feedback all required reflection, application, and personalisation of strategies, which are processes known to help people internalise new skills and promote behaviour change [66] [67].

Interestingly, the reduction in stress scores was also seen in the control group. This is likely to reflect the effect of drawing participants' attention to digital stress through the definitions and questionnaires, together with the time available between assessments. Simply becoming aware of the different forms of digital stress and having an opportunity to reflect on them over several weeks, may have encouraged small adjustments in everyday digital habits even without formal intervention. For example, a related phenomenon known as the mere measurement effect has been widely documented in consumer psychology. Simply asking people about their intentions before making a purchase can influence their decision-making, as it increases access to their own attitudes and makes the most salient option in a set of choices more likely to be selected [68].

Contrary to expectations, the results revealed no significant differences between the groups in the reduction of digital stress. Does the absence of differences between groups mean that interventions are redundant? In this study, all participants, including those in the control condition, were exposed to definitions of digital stress and completed detailed questionnaires at two time points. This process alone may have increased awareness of availability demands, approval anxiety, FoMO, connection overload, and online vigilance, encouraging small changes over the following weeks. Such reactivity to measurement is well documented in behavioural research [69][70]. By contrast, the interventions were carefully designed, with structured practice, scenario-strategy work, feedback, and opportunities for reflection. Yet, we did not detect the differences between groups in reducing digital stress or its component. This raises important questions for future research. For example, are the awareness-driven reductions observed here transient, or can they sustain change over longer periods? Evidence from the digital domain suggests that awareness-only effects can be short-lived or modest. For example, providing people with usage estimates for two weeks reduced perceived stress from social networking without changing actual use (i.e., awareness shifted stress appraisals rather than behaviour), whereas two-month pop-up notifications about "excessive use" did not reduce screen time or problematic smartphone use, and screen-time tracking increased self-awareness but was unlikely to reduce usage [71]. Perhaps, a much longer post-intervention assessment in interventions designed to reduce digital stress can shed light on the time-related dynamics of the intervention effects.

Other questions that are prompted by the absence of between-group differences in our study are: would sharper contrasts between types of structured engagement reveal unique mechanisms, such as the role of accountability, teaching expectancy, or active search in reinforcing learning and coping strategies in reducing digital stress? And to what extent do different components of digital stress, such as approval anxiety, FoMO, or online vigilance, respond preferentially to specific forms of structured practice rather than to awareness alone? By demonstrating robust improvements across all groups in reducing digital stress a rigorously controlled design, the present study lays the foundation for future research to answer these questions.

From a different angle of consideration, although protégé effect can enhance learning and cognitive processing[15], modifying online behavior such as constant checking, social comparison, and FoMO, may require more than reflective learning alone. These behaviors are often maintained by a combination of habitual routines, persuasive design features [72], personal factors, and pervasive social expectations and norms [73]. Therefore, changing or building resistance to these patterns may require multifaceted interventions beyond reflection.

The significant within-subject main effect on overall digital stress and its components (availability demand stress, approval anxiety, FoMO, connection overload, and online vigilance) indicates a general trend of reduced digital stress regardless of the group. However, the lack of a significant interaction effect between time and group for these measures suggest that the protégé-based interventions (passive and active learning) did not yield more reduction in digital stress compared to the digital literacy or control groups. This finding challenges the study hypothesis that engaging in a teach-to-learn would enhance reflection and internalization of coping strategies. It is possible that the mechanisms underlying digital stress reduction, particularly those stem from social expectations and norms, are resistant to short-term change.

The findings may also be interpreted through the lens of culture, particularly the participants' collectivist background (e.g. Middle Eastern cultures). In collectivist societies, individuals tend to prioritize group harmony and conformity over personal preferences [74]. When social norms emphasize staying connected and being responsive to others, individuals may feel obligated to comply, even when such digital availability causes stress. Moreover, studies have shown that norm violators in collectivist contexts face social disapproval[75]. In this context, the stress of violating group expectations may be more psychologically stressful than the stress of conforming, thereby limiting the effectiveness of reflective strategies.

Another possible explanation for the absence of intervention effects could be due to the limited behavioural engagement of the teaching task. Although preparing to teach is cognitively engaging, it may not be sufficiently to drive behavioural change. In the current design, the protégé-based intervention was implemented through a hypothetical teaching task without actual delivery to peers, which may have limited participants perceived sense of responsibility and social obligation to internalize and act on the coping strategies. While the use of learner characters and simulated feedback aimed to increase realism and personal relevance, an approach aligned with prior studies using teachable agents to invoke responsibility [50], actual engagement may have enhanced the outcome. Prior research has suggested that the interactive teaching, where learners interact with peers and receive to real time feedback, may increase the motivation and effect of learning[76]. Thus, while reflective preparation may initiate awareness of digital stress, online behaviour change may require the added reinforcement of social presence.

Given that digital stress is related to PSMU and has been shown to mediate the relationship between PSMU and adverse outcomes [20] [21], it was expected that reduction in digitals stress would yield to corresponding reduction in PSMU. However, as no significant reductions in digital stress were observed between the groups, no changes in PSMU were observed either. Additionally, secondary benefits in reducing PSMU were hypothesized based on the concept of cross-protection, as applied in inoculation theory-based studies, which posits that resistance in one domain may confer protection in related domains [77]. Since digital stress levels did not significantly differ between groups, the lack of reduction in PSMU aligns with this concept.

Similarly, issue involvement in managing digital stress did not show a significant increase in this study. One possible explanation is that the intervention did not stimulate sufficiently deep reflection or enhance motivation beyond participants' existing levels of motivation. This interpretation is supported by the high baseline scores of issue involvement across all groups, suggesting that participants already viewed managing digital stress as a highly relevant and meaningful topic. While high personal relevance is often considered a prerequisite for cognitive engagement and attitude change[78], our findings indicate that it may not be sufficient on its own to enhance the effectiveness of behavioral interventions. Additionally, no significant differences were observed for positive talk with others (WOM) about managing digital stress. This suggests that the intervention did not prompt participants to discuss or advocate for digital stress management strategies. One interpretation is that when an issue is already perceived as relevant and important, the common drivers of WOM, such as the need for reassurance or advocacy may be less activated [79]. Participants may have already formed confident positions on the topic, not increasing their motivation to engage in discussion with others. Another possibility is that the process of preparing teaching materials in the protégé-based groups served as a substitute channel for self-expression or social interaction, which are motives of WOM [80], satisfying the urge to articulate their views and thereby reducing the subsequent need to engage in WOM.

The findings have several implications for the technology design and implementation of interventions aimed at mitigating digital stress. Firstly, the lack of significant differences between the intervention groups highlights the complexity of changing ingrained digital habits and stressors. It could be that cognitive or reflective engagement may not be sufficient on its own to change online behaviors that are reinforced by different factors including habit and technology design. This aligns with behaviour change research suggesting that increased knowledge or awareness should be paired with changes in environment or habit cues to achieve behavioral impact [81] [82].

Digital stress is influenced by a combination of personal factors, the persuasive design of social media platforms, and social pressures. This multifaceted nature may suggest that a personalized intervention that consider an individual's specific stressor, personality traits, and social context may be more effective. For instance, interventions could be tailored to address the specific dimensions of digital stress that are most prominent for an individual, such as approval anxiety or FoMO. For example, the FoMO-R intervention, which used a diary-based method to promote self-reflection on social media use, demonstrated a reduction in FoMO. This approach could be integrated with protégé-based intervention to address other digital stressors rooted in social expectations [83]. Additionally, given the influence of social pressure and norms on digital stress, especially in collectivist cultural contexts, future interventions could combine protégé-based intervention with components specifically designed to reduce perceived social pressure. For example, a systematic review[84] found that interventions incorporating social-norm strategies, such as peer discussions and online support forums, were effective in promoting health-related behaviors, such as increasing reducing heavy or risky drinking. Integrating such components may help individuals not only reflect on stress-inducing norms but also feel socially supported in resisting them.

From an HCI and design perspective, future designs could incorporate teachable agents or peer forums where users actually share advice and experiences, thereby invoking a stronger sense of responsibility and commitment. Prior work on the protégé effect with teachable agents has shown that learners put more effort when they feel responsible for teaching an agent or another person [50]. Furthermore, recent research highlights that online experiences are multifaceted [85], with social media can perceived as both a contributor to problematic use and a facilitator of social well-being [86]. Therefore, interventions and social media design should aim to balance these opposing effects, reducing digital stress stemming from social pressure while preserving the positive social functions that support users' social well-being. This aligns with research suggesting for digital platforms to promote healthier digital use by strengthening users' wellbeing skills such as awareness and purpose [87].

This study has a number of limitations that should be considered when interpreting the findings. The study relied on self-report measures for digital stress and PSMU. Future studies could incorporate more objective measures, such as data from smartphone usage tracking. Second, the study implemented a protégé-based intervention using a hypothetical teaching task. While the "learning by teaching" principle is well-established, the absence of real-time interaction and feedback may have limited the intervention's impact. Although this limitation was mitigated by simulating learner personas and providing feedback framed as learner questions, future studies could explore live or interactive teaching formats (e.g., teaching peers or virtual agents) to enhance participants' sense of social accountability and engagement. Finally, the study's sample was recruited from a specific cultural context (Middle Eastern countries), where collectivist values tend to emphasize conformity, group harmony, and sensitivity to social expectations [88]. In such contexts, individuals may be more inclined to internalize digital availability norms or approval-seeking behaviors as part of maintaining group cohesion. This cultural orientation could influence both the baseline experience of digital stress and how individuals respond to interventions aimed at reducing it. Therefore, the generalizability of the findings to individualistic cultures, where autonomy and self-expression are prioritized, may be limited. Further research is needed to explore how the experience of digital stress and the effectiveness of interventions may vary across different cultural contexts.

## 5 Conclusion

This study evaluated a socio-technical intervention based on the protégé effect in managing digital stress and its components, as well as its influence on PSMU, issue involvement, and WOM. While prior literature has supported the cognitive and motivational benefits of the protégé effect in educational contexts, our findings suggest that its application to digital stress reduction and behavioural change in online environments requires further investigation.

This study contributes to the growing body of research on digital stress and interventions designed to promote digital wellbeing. While the protégé-based intervention did not demonstrate more effect compared to digital literacy or a control condition, the findings provide valuable insights for future research and practice. Future studies may need to incorporate more interactive, and personalized interventions that go beyond raising awareness to actively engage individuals in behavioural change. As digital technologies increasingly integrate into our daily lives all aspects of our daily lives, the development of effective strategies for managing digital stress will become increasingly critical for promoting mental health and wellbeing in the digital age.

## Acknowledgement

This publication was supported by NPRP 14 Cluster grant # NPRP 14C-0916–210015 from the Qatar National Research Fund (a member of Qatar Foundation). The findings herein reflect the work and are solely the responsibility of the authors.

## Supplementary material

The study design and dataset are available on the Open Science Framework (OSF) at the following link: https://osf.io/kqs54/?view_only=ed0f6660c5b740dfbcc3ad415943c445

## Authors Contribution

SA: Conceptualised the research, designed the study, curated the Data, designed and performed and reported the statistical analysis, wrote the first draft. AY: Designed the analysis methodology, validated the statistical analysis, reviewed and edited the paper. DA: Conceptualised the study, reviewed and revised the study design, reviewed and edited the paper. RA: Conceptualised the research, designed the study, validated the analysis, reviewed and edited the paper.